\documentclass[aip,sd, amsmath,amssymb,floatfix,reprint]{revtex4-1}

\usepackage{graphicx}
\usepackage{dcolumn}
\usepackage{bm}

\begin{document}

\title{Defect calculations in semiconductors through a dielectric-dependent hybrid DFT functional: the case of oxygen vacancies in metal oxides}

\author{Matteo Gerosa}
\affiliation{Department of Energy, Politecnico di Milano, via Ponzio 34/3, 20133 Milano, Italy}
\author{Carlo Enrico Bottani}
\email[Corresponding author: ]{carlo.bottani@polimi.it}
\affiliation{Department of Energy, Politecnico di Milano, via Ponzio 34/3, 20133 Milano, Italy}
\affiliation{Center for Nano Science and Technology @Polimi, Istituto Italiano di Tecnologia, via Pascoli 70/3, 20133 Milano, Italy}
\author{Lucia Caramella}
\author{Giovanni Onida}
\affiliation{Dipartimento di Fisica, Universit\`a degli Studi di Milano, Via Celoria 16, 20133 Milano, Italy}
\affiliation{European Theoretical Spectroscopy Facility (ETSF)}
\author{Cristiana Di Valentin}
\author{Gianfranco Pacchioni}
\affiliation{Dipartimento di Scienza dei Materiali, Universit\`a di Milano-Bicocca, via R. Cozzi 55, 20125 Milan, Italy}

\date{\today}

\begin{abstract}
 We investigate the behavior of oxygen vacancies in three different metal-oxide semiconductors (rutile and anatase TiO$_2$, monoclinic WO$_3$, and tetragonal ZrO$_2$) using a recently proposed hybrid density-functional method in which the fraction of exact exchange is material-dependent but obtained \emph{ab initio} in a self-consistent scheme. In particular, we calculate charge-transition levels relative to the oxygen-vacancy defect and compare computed optical and thermal excitation/emission energies with the available experimental results, shedding light on the underlying excitation mechanisms and related materials properties. We find that this novel approach is able to reproduce not only ground-state properties and band structures of perfect bulk oxide materials, but also provides results consistent with the optical and electrical behavior observed in the corresponding substoichiometric defective systems.
\end{abstract}

\maketitle

\section{Introduction}

Oxygen vacancies represent a prototypical intrinsic point defect in metal-oxide materials, where they are inevitably formed throughout the synthesis process, as a result of the exposition of the samples to high temperatures and strongly reducing atmospheres. 

The presence of oxygen vacancies substantially alters the optical and electrical properties of metal-oxide semiconductors,\cite{ganduglia2007,pacchioni2003} and a correct interpretation of experiments requires these modifications to be properly taken into account. Moreover, defects control and engineering play a fundamental role in a wide range of applications, such as in catalysis and photocatalysis,\cite{pacchioni2015a,divalentin2013} information technology,\cite{sawa2008} gas sensors\cite{krotcenkov2007} and smart windows.\cite{niklasson2007}

The standard approach to point defects computational modeling in materials is rooted within density-functional theory (DFT) and the supercell approximation.\cite{freysoldt2014} The simplest interpretation of the spectroscopic features of defective materials may be provided by analysis of the corresponding Kohn-Sham (KS) electronic band structures. However, it is well known that KS energy eigenvalues represent a very crude approximation to quasiparticle energies probed in spectroscopies. The limits of such approximation noticeably manifest themselves in the failure of standard DFT in predicting accurate electronic band gaps. An exact xc functional, providing exact total energies, would be able to yield exact defect states through comparison of defect formation energies corresponding to variations of the charge state of the defect (\emph{charge-transition levels}, CTLs).\cite{scherz1993,vandewalle2004,lany2008} In principle, CTLs can be rigorously put in correspondence with experimentally measurable excitation and emission energies in defective semiconductors and insulators;\cite{gallino2010} in an analogous spirit, the delta-self-consistent-field method allows one to compute electronic excitations in finite systems from total-energy differences.\cite{onida2002} In practice, of course, the error due to an approximate form of the xc term would affect total energies and hence the computed CTLs. Notice in fact that standard local/semilocal DFT functionals often predict defect states to be resonant with the bulk band edges, even when experiments indicate a deep donor or acceptor behavior; in this case, computed total energies, and the corresponding CTLs, would be largely incorrect.\cite{deak2011pssb}
Hence, improving the functional with respect to local/semilocal ones is a prerequisite for the application of the method. Promising results have been achieved by adopting exact-exchange hybrid DFT functionals\cite{becke1993a} for the description of exchange and correlation.\cite{divalentin2014} Such functionals have been proven to yield band gaps in more quantitative agreement with experiments, as well as to help correcting for the self-interaction error inherent in local/semilocal approximations.  

In the frame of hybrid DFT calculations on extended systems, a connection between the fraction of exact exchange to be admixed in the hybrid functional and the dielectric constant of the material has been proposed.\cite{alkauskas2011,marques2011} The obtained functional has proved capable of predicting band gaps, structural properties and total energies for semiconductors in quantitative agreement with experiments.\cite{marques2011,skone2014,gerosa2015} In particular, in our recent work\cite{gerosa2015} we tested the performance of this approach in calculating various properties of metal-oxide semiconductors.

Motivated by this success, we devote the present work to an assessment of the efficiency of our method in describing defective oxide semiconductors, with particular concern to oxygen vacancies, which constitute a prototypical case study in this kind of materials. In this respect, we investigate the properties of three different oxide semiconductors of great both fundamental and applicative interest: rutile and anatase titanium dioxide (TiO$_2$), tungsten trioxide (WO$_3$) in its room-temperature (RT) $\gamma$-monoclinic crystal structure, and tetragonal zirconium dioxide (ZrO$_2$).  

The paper is organized as follows. In Section~\ref{sec:methods}, details on the performed DFT calculations are given, the dielectric-dependent hybrid DFT method is discussed and the charge-transition levels formalism is reviewed. In Section~\ref{sec:results}, results are discussed and interpreted in the light of the available experiments on reduced oxide samples. Finally, the most important conclusions are summarized in Section~\ref{sec:conclusions}.

\section{Computational approach}
\label{sec:methods}

\subsection{DFT calculations}

All the calculations were performed using the \textsc{crystal09} code,\cite{dovesi2005,crystal09-manual} based on linear combination of atomic orbitals and Gaussian-type basis sets. Adoption of small-sized localized atomic-like functions for representing KS orbitals (compared, e.g., to much larger plane-wave basis sets) allows one to considerably reduce the computational cost of hybrid-functional calculations almost down to that of standard local/semilocal DFT calculations. This in turn allows to employ moderately large supercells (containing roughly a hundred atoms) for modeling defective materials, requiring a reasonable computational effort also when performing structural optimizations.

In the case of TiO$_2$, calculations were performed in the all-electrons scheme, using the basis set from Ref.~\onlinecite{wilson1998} for Ti. Small-core effective-core pseudopotentials were employed for modeling core electrons for W and Zr atoms\cite{hay1985} in WO$_3$ and ZrO$_2$, while valence electrons were described using the basis sets defined in Ref.~\onlinecite{wang2011} ($5p$, $5d$, $6sp$ W electrons in the valence) and Ref.~\onlinecite{bredow2004} ($4d$, $5sp$ Zr electrons in the valence), respectively. The O atom was always treated at the all-electron level, using the basis set from Ref.~\onlinecite{ruiz2003} for TiO$_2$ and WO$_3$, and from Ref.~\onlinecite{gallino2011} for ZrO$_2$.

We considered bulk 72-atom $2\times2\times3$ and 96-atom $2\sqrt{2}\times2\sqrt{2}\times2$ supercells for rutile and anatase TiO$_2$, respectively. For RT $\gamma$-monoclinic WO$_3$, we adopted a model supercell comprising 64 atoms, obtained by doubling the primitive cell along the $a$ crystallographic axis. For tetragonal ZrO$_2$, a 108-atom $2\times2\times3$ supercell was employed. The O vacancy was modeled by removing one O atom from the corresponding supercells. The Brillouin zone was sampled using $2\times2\times2$ $\Gamma$-centered Monkhorst-Pack grids,\cite{monkhorst1976} corresponding to 6 or 8 $k$ points in the irreducible wedge.

Structural optimizations were performed by allowing all the atoms in the cell to relax their positions, keeping the lattice parameters fixed to that of the optimized bulk cell. Convergence thresholds in geometry optimizations were set at their standard values in the \textsc{crystal09} code: the defined thresholds for the maximum and the root-mean-square of the energy gradients (atomic displacements) are $0.00045$~a.u. ($0.00180$~a.u.) and $0.00030$~a.u. ($0.00120$~a.u.), respectively.\cite{crystal09-manual}

\subsection{Dielectric-dependent self-consistent hybrid functional}

The xc potential was approximated within the generalized KS scheme,\cite{seidl1996} using a full-range hybrid functional\cite{becke1993a} in which the fraction of exact exchange $\alpha$ is computed self-consistently with the static electronic dielectric constant $\epsilon_{\infty}$.\cite{alkauskas2011,skone2014,gerosa2015} In the framework of many-body perturbation theory (and in particular within the $GW$ scheme), it can be proven that $\alpha\approx 1/\epsilon_{\infty}$, provided that microscopic components of the dielectric function are averaged out over the cell and dynamical effects are neglected (in which case the $GW$ self-energy reduces to the Hedin's Coulomb-hole-plus-screened-exchange approximation\cite{hedin1965}).

The electronic dielectric constant $\epsilon_{\infty}$ was computed within first-order perturbation theory using the coupled-perturbed Kohn-Sham approach implemented in \textsc{crystal09},\cite{ferrero2008} allowing for evaluation of the xc contribution to electronic screening beyond the random-phase approximation.

The dielectric-dependent self-consistent hybrid functional was defined for the bulk pristine crystal, and then applied to the corresponding defective system. This procedure is justified by the facts that (i) the method proved able to provide electronic gaps, structural properties and ground-state energies in quantitative agreement with experiment for bulk systems,\cite{marques2011,skone2014,gerosa2015} and (ii) the presence of point defects is expected to negligibly affect the dielectric properties in the dilute limit of defect concentration which is ultimately addressed.

The self-consistent procedure yielded the following values for the exchange fraction: $15.2\%$ for rutile and $18.4\%$ for anatase TiO$_2$, $21.9\%$ for RT monoclinic WO$_3$, $20.8\%$ for tetragonal ZrO$_2$. Notice that these values are close to, but do not exactly match, those entering the definition of standard hybrid functionals, such as PBE0 ($25\%$) and B3LYP ($20\%$).\cite{perdew1996,stephens1994,becke1993b} Also, the corresponding dielectric constants are found close to the experimental values, deviating from them by at most four percent (see also Refs.~\onlinecite{gerosa2015,skone2014}). 

\subsection{Charge-transition levels}

The position of the electronic levels introduced by O vacancies in the band gap of the material was computed on the basis of the CTLs formalism.\cite{scherz1993,lany2008,vandewalle2004,freysoldt2014} The latter constitutes a rigorous framework for computing excitation and emission energies in defective semiconductors that can be directly compared with experiments (see the Introduction).

At fixed experimental conditions, corresponding for instance to exposing the material to high-temperature O-rich or O-poor atmospheres in the case of oxide-based materials, the formation energy of the defect (O vacancy) depends only on the chemical potential (which at $T=0$ and without an external field can be identified with the Fermi energy) of the electrons that can be exchanged between the defect and the host crystal subsystems, following some external (e.g., optical) perturbation.

The optical transition level $\mu^{\text{opt}} (q/q')$ is defined as the Fermi energy at which the formation energies of the defect in the (dimensionless) charge states $q$, $E_{\text D,q}^{\text{f}}$, and $q'$, $E_{\text{D},q'}^{\text{f}}$, become equal. Since we are dealing with optical (vertical) excitations, occurring over time-scales much shorter than the phonon time-scale, the atomic configuration is kept frozen at that of the initial defect charge state, and all the quantities of interest (e.g., the formation energy) are evaluated at that configuration for both charge states.\cite{note1} The optical transition level, referred to the top of the valence band (VB), $\varepsilon_{\text v}$, of the host crystal, can thus be expressed as
\begin{equation}
 \mu^{\text{opt}} (q/q') = \frac{E_{\text D,q'}-E_{\text D,q}}{q-q'} - \varepsilon_{\text v},
 \label{eq:tl}
\end{equation}
where the the ground-state total energies of the defective supercells with defect in the charge states $q$ and $q'$, $E_{\text D,q}$ and $E_{\text D,q'}$, are presently assumed to incorporate any corrections due to electrostatic interactions between charged image defects and to the definition of a common energy reference for the two bulk calculations.\cite{komsa2012}

Using Janak's theorem,\cite{janak1978} and in the spirit of Slater's transition-state theory,\cite{slater1972} the total energy difference appearing in Eq.~\eqref{eq:tl} can be approximated as\cite{alkauskas2007,gallino2010}
\begin{equation}
 E_{\text D,q'}-E_{\text D,q} \approx \frac{1}{2}\left[\varepsilon_N (N;q') + \varepsilon_{N}(N-1;q=q'+1)\right],
 \label{eq:janak-slater}
\end{equation}
where $\varepsilon_N (N;q')$ is the KS eigenvalue of the highest occupied spin-orbital of the $N$-electron system (corresponding to defect charge state $q'$) and $\varepsilon_N (N-1;q=q'+1)$ is the KS eigenvalue of the lowest unoccupied spin-orbital of the $(N-1)$-electron system, corresponding, for instance, to excitation of an excess electron into the host crystal (defect charge state $q=q'+1$).\cite{note2}

Two issues have to be considered in the evaluation of the KS eigenvalues in Eq.~\eqref{eq:janak-slater}:
\begin{itemize}
 \item The KS eigenvalues obtained from the two bulk calculations (defective supercells corresponding to different charge states $q$ and $q'$) have to be referenced to some common energy scale. We chose the zero of the energy to be the $1s$ KS eigenvalue of O (located $\sim 500$~eV below the VB edge), averaged over the whole supercell.
 \item For charged defects, the electrostatic interactions between image defects arising from usage of finite-size supercells have to be taken into account.\cite{lany2008,komsa2012} The corresponding correction to the KS defect level $\varepsilon_{\text D}$ reads\cite{chen2013}
 \begin{equation}
  \Delta \varepsilon_{\text D}(N;q) = -\frac{2}{q} \Delta E_{\text D,q}^{\text{MP}} = -2 (1+f) \left(\frac{\alpha_{\text M}}{2L}\right) \frac{q}{\epsilon_{\infty}},
 \end{equation}
where the correction $\Delta E_{\text D,q}^{\text{MP}}$ to the total energy of the charged supercell is evaluated in the Makov-Payne scheme;\cite{leslie1985,makov1995} $\alpha_{\text M}$ is the Madelung constant, $L$ is the average distance between image defects, and  $\epsilon_{\infty}$ is the static electronic dielectric constant of the host crystal, computed within the self-consistent hybrid scheme.\cite{note3} The prefactor $f\approx-0.35$ approximately takes into account the interaction energy of the defect charge with the neutralizing jellium background (scaling as $L^{-3}$) introduced to renormalize total energy in charged defect calculations.\cite{lany2008}
\end{itemize}

Thermodynamic levels $\mu^{\text{therm}}(q/q')$, corresponding to excitations $q_i \rightarrow q_f$ in which the system is allowed to relax in the final charge state $q_f$ (adiabatic transitions) are obtained from optical levels by adding or subtracting the relaxation energy\cite{gallino2010}
\begin{equation}
 E_{\text{rel}}=\left.E_{\text D,q_f}\right|_{q_i\,\text{geom}}-\left.E_{\text D,q_f}\right|_{q_f\,\text{geom}},
 \label{eq:rel-energy}
\end{equation}
 according to whether $q\equiv q_f=q_i+1\equiv q'+1$ (electron excitation) or $q'\equiv q_f=q_i-1 \equiv q-1$ (electron trapping).\cite{gallino2010} Notice that in Eq.~\eqref{eq:rel-energy} electrostatic corrections to formation energies would have to be evaluated for the same charge state $q_f$, and they thus cancel against each other.\cite{deak2011}

\section{Results and discussion}
\label{sec:results}

\subsection{TiO$_2$}

The optoelectronic properties of O-deficient TiO$_2$ (TiO$_{2-x}$) have been extensively investigated both experimentally\cite{diebold2003} and theoretically,\cite{finazzi2008, janotti2010, mattioli2010, mattioli2008, deak2012, malashevich2014, divalentin2014} making this system an ideal model for testing the performance of the dielectric-dependent hybrid method in describing the electronic properties of defective semiconductors. 

The observed $n$-type conductivity of TiO$_2$ samples has been related to the presence of O vacancies, acting as intrinsic donors, formed under strongly reducing synthesis conditions.\cite{diebold2003} The larger intrinsic conductivity of the anatase phase, which also makes it more appealing for applications, has been explained with the different behavior of the O vacancy in the two polymorphs.\cite{mattioli2010,deak2012}

On the other hand, optical experiments found the associated defect levels to be located deep in the band gap. Infrared absorption (IR),\cite{cronemeyer1959} ultraviolet photoelectron,\cite{nolan2008} and electron energy-loss experiments\cite{henderson2003} on reduced rutile samples all agree in attributing a feature at about 1~eV below the conduction band (CB) to the presence of O vacancies. Theoretical calculations have revealed that the apparent contradiction between the observed $n$-type conductivity and the deep nature of optical levels in TiO$_{2-x}$ is rationalized by the different character (shallow versus deep) of the O-vacancy thermodynamic and optical transition levels.\cite{deak2012, mattioli2010}

Furthermore, the excess charge due to O deficiency has been shown to redistribute differently in the host crystal, according to whether the rutile or anatase polymorph is considered, resulting in the localization of the extra electrons on topologically different Ti atoms.\cite{setvin2014}

It is the goal of the present part of the work to re-investigate these issues within the dielectric-dependent hybrid approach, which by construction is sensitive to the different electronic screening characteristic of the rutile and anatase phases.

\subsubsection{Rutile}
\label{sec:rutile}

\begin{figure}[tb]
 \includegraphics{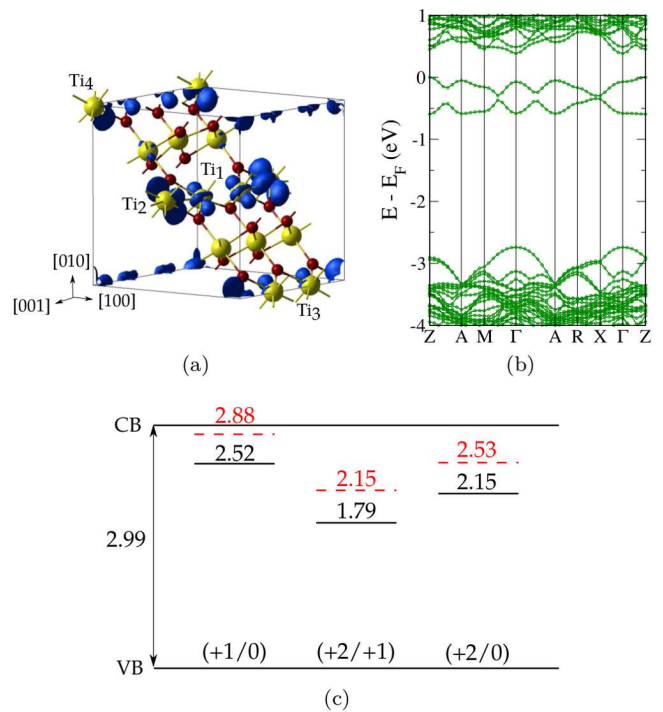}
 \caption{O vacancy in rutile TiO$_2$. (a) Isosurface plot of the charge density associated with the excess electrons for the neutral vacancy. The yellow (large) and red (small) spheres represent Ti and O atoms, respectively. Only the atomic plane containing the O vacancy is shown. (b) Corresponding electronic band structure (only spin majority is shown) computed at the equilibrium geometry. (c) Optical (solid, black) and thermodynamic (dashed, red) charge-transition levels. Positions (in eV) are given with respect to the top of the valence band (VB).}
 \label{fig_1}
\end{figure}

The ground state of rutile TiO$_2$ with a neutral O vacancy is found to be a triplet, with the two excess electrons redistributing over the whole (110) plane containing the vacancy [Figure~\ref{fig_1}(a)]. Most of the excess charge is contributed by the fully-coordinated Ti$_2$ (spin density $n_s = n_{\uparrow} - n_{\downarrow}=0.70$), Ti$_3$ (two equivalent atoms, $n_s=0.15$ each) and Ti$_4$ ($n_s = 0.65$) atoms. The two under-coordinated Ti$_1$ atoms instead accommodate a smaller fraction of it ($n_s=0.13$ each). This picture is in qualitative agreement with recent scanning tunneling microscopy (STM) measurements on the rutile (110) surface,\cite{setvin2014} suggesting the localization of the extra electrons on Ti atoms away from the vacancy,\cite{note4} with the formation of a polaron. Electron paramagnetic resonance (EPR) measurements also confirmed the presence of Ti$^{3+}$ ionic species formed as a consequence of electron trapping at Ti$^{4+}$ centers in TiO$_{2-x}$ crystals.\cite{yang2009,brandao2009}

Notice that the localization of the unpaired electrons at Ti$^{3+}$ centers is correctly described only within the hybrid-functional scheme,\cite{divalentin2006} at variance with the case of standard local/semilocal functionals which commonly fail, due to incomplete cancellation of the self-interaction error.   
Notice also that the above physical picture is correctly captured only at the level of spin-polarized DFT, while neglecting spin polarization leads to a completely different description, with the excess charge trapped in the vacancy void,\cite{janotti2010, malashevich2014} in a configuration typical of a color center (F center). We found the corresponding closed-shell ground state to be highly unstable, being above the triplet solution by nearly 2~eV in total energy. In the spin-compensated case, the two under-coordinated Ti atoms relax towards the vacancy, shortening their distance to $2.90\,\text{\AA}$ (compared with an unrelaxed distance of $2.99\,\text{\AA}$); instead, in the true, spin-polarized ground state (triplet), they move away from the vacancy, resulting in an equilibrium distance of $3.37\,\text{\AA}$.

Our found magnetic ground state is compatible with the evidence of triplet and doublet signals obtained in EPR measurements\cite{yang2009, brandao2009} on rutile TiO$_{2-x}$ crystals, and attributed to neutral and singly-charged O vacancies, respectively.

We succeeded in stabilizing a triplet configuration in which nonetheless the excess charge is confined at the vacancy void,\cite{note5} a situation typically encountered in non-reducible O-deficient oxides.\cite{pacchioni2015} However, we found the corresponding ground-state energy to be $\sim 0.8$~eV higher than that of the triplet polaron-like solution represented in Figure~\ref{fig_1}(a). This finding definitely rules out the possibility of an F-center-like behavior of O vacancies in rutile, in favor of the polaron picture, which is in agreement with all the available experimental evidence. This conclusion is also corroborated by previous theoretical investigations using larger supercells.\cite{deak2012}

The bulk band gap of rutile TiO$_2$, as computed using the self-consistent hybrid functional, is found to be 2.99~eV, a value compatible with the available data from photoemission/inverse photoemission experiments, yielding an electronic gap of $3.3 \pm 0.5$~eV (Ref.~\onlinecite{tezuka1994}) and $\sim 3.1$~eV (Ref.~\onlinecite{see1994}). Upon removal of an O atom, the two resulting excess electrons occupy two triplet defect states appearing in the band gap at about $0.5-1$~eV below the CB [Figure~\ref{fig_1}(b)]. As the vacancy is ionized, the remaining excess electron localizes on the three Ti atoms in the row of the O vacancy, again with a dominant contribution from the fully-coordinated Ti$_2$ atom ($n_s = 0.68$). The associated occupied defect level is still localized at about 1~eV below the CB.

Figure~\ref{fig_1}(c) shows the computed charge-transition levels. The optical levels lie deep in the band gap, resulting in excitation energies from the defect states to the CB of 0.47~eV, 1.20~eV and 0.84~eV for the $(+1/0)$, $(+2/+1)$ and $(+2/0)$ transitions, respectively. Notice that, as a result of the finite supercell size, the residual defect-defect interaction (i.e. overlap between polarons)\cite{sezen2014} leads to a small dispersion of the highest-occupied defect band (about 0.2~eV). This should be considered as indicative of the numerical accuracy of the computed level. Being aware of this limitation, our results can be considered consistent with the features, found at about 1~eV below the bottom of the CB, observed in experimental spectra.\cite{cronemeyer1959,henderson2003,nolan2008} In particular, IR measurements\cite{cronemeyer1959} revealed two absorption peaks at 0.75~eV and 1.18~eV for the single and double ionization, which correlate well with our computed first and second optical ionization energies for the O vacancy.\cite{note7}

Thermodynamic transition levels are significantly higher in energy, reflecting the sizable structural relaxation associated with the polaronic distortion. The $(+1/0)$ adiabatic transition is located about 0.1~eV below the CB, indicating that the neutral O vacancy may be stable at sufficiently low temperature. The second transition $(+2/+1)$ is at 0.81~eV, while the double ionization [$(+2/0)$ transition] requires 0.46~eV. These values should be compared with measured thermal ionization energies observed in a range between 0.3 and 0.6~eV.\cite{ghosh1969}

\subsubsection{Anatase}

The nature of the neutral O vacancy in anatase TiO$_2$ has been extensively studied in the theoretical literature.\cite{finazzi2008,deak2012,mattioli2010,mattioli2008,morgan2010,divalentin2009,deangelis2014} However, no consensus has been achieved so far concerning the redistribution of the associated excess electrons in the host crystal. This difficulty may be rationalized with the presence of several minima on the relevant adiabatic potential energy surface. The realization of a specific configuration is strongly related to the description of the structural relaxation around the vacancy, which is affected, for instance, by the employed DFT xc approximation.

\begin{figure}[tb]
  \includegraphics{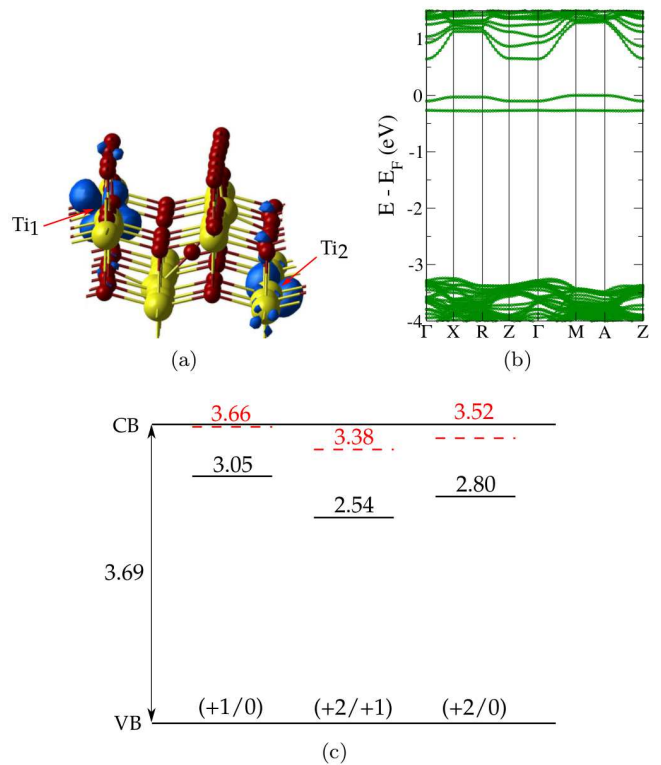}
 \caption{O vacancy in anatase TiO$_2$. (a) Isosurface plot of the charge density associated with the excess electrons for the neutral vacancy. The yellow (large) and red (small) spheres represent Ti and O atoms, respectively. (b) Corresponding electronic band structure (only spin majority is shown) computed at the equilibrium geometry. (c) Optical (solid, black) and thermodynamic (dashed, red) charge-transition levels. Positions (in eV) are given with respect to the top of the VB.}
 \label{fig_2}
\end{figure}

Our calculations yield a triplet ground state for the system with a neutral O vacancy, with a corresponding substantial rearrangement of the atomic positions around the vacancy. As shown in Figure~\ref{fig_2}(a), this implies a reduction of the local symmetry around the vacancy, with a nearest-neighbor O atom breaking its bond with a bulk Ti atom (Ti$_2$) and considerably relaxing towards the vacancy void.\cite{note6} The same behavior was observed in all previous calculations in which symmetry-breaking relaxation was allowed for.\cite{mattioli2008,finazzi2008,morgan2010} The two excess electrons are found to be localized on two Ti atoms (the under-coordinated Ti$_1$, with $n_s = 0.92$, and the bulk Ti$_2$, with $n_s = 0.84$), similarly to a previous DFT+U investigation.\cite{morgan2010} This finding is again in qualitative agreement with the STM investigation of Diebold and coworkers,\cite{setvin2014} which suggests that the excess charge preferably stays close to the vacancy in anatase, as opposed to the case of rutile in which the major contribution is given by the fully-coordinated Ti atoms (see Section~\ref{sec:rutile}).

For the system with a neutral vacancy, the computed electronic structure shows two triplet defect states at $\sim 0.7-1$~eV below the CB [Figure~\ref{fig_2}(b)]. The computed band gap is 3.69~eV, larger than the value of 3.42~eV obtained from optical absorption experiments at low temperature.\cite{tang1995} To the best of the authors' knowledge, no data for the photoemission gap of stoichiometric anatase TiO$_2$ are available in the experimental literature. However, our computed gap is consistent with previous $GW$ calculations.\cite{landmann2012,kang2010,gerosa2015}

When the vacancy is ionized, a doublet occupied defect state locates at $\sim 0.7$~eV below the CB. The extra electron remains localized ($n_s = 0.92$) at the under-coordinated Ti$_1$ atom.

The computed thermodynamic transition levels are significantly shallower than in rutile [compare Figure~\ref{fig_1}(c) and Figure~\ref{fig_2}(c)]. In particular, the $(+1/0)$ level is practically resonant with the CB, suggesting that an electron can be thermally excited into the CB, thus accounting for the observed larger conductivity of anatase samples.

Similarly to the case of rutile, optical levels are substantially deeper than the corresponding thermodynamic ones. We computed excitation energies of 0.64~eV, 1.15~eV and 0.89~eV for the $(+1/0)$, $(+2/+1)$ and $(+2/0)$ transitions. Experimentally, a feature at $\sim 1.0-1.1$~eV below the CB has been reported in different experiments (combined x-ray photoemission/absorption spectroscopy,\cite{thomas2007} as well as scanning tunneling spectroscopy\cite{setvin2014}) on reduced anatase surfaces, and has been attributed to O vacancies. Notice that, since the stable charge state of the O vacancy is $q=+1$ at room temperature, it seems likely that the $(+2/+1)$ transition is mainly probed in optical experiments; accordingly, our computed excitation energy of 1.15~eV matches excellently with the measured spectroscopic features.

In conclusion, the present approach proved capable of explaining several experimental signatures related to the different behavior of O vacancies in the two most common TiO$_2$ polymorphs, rutile and anatase. In particular, the larger conductivity of anatase can be understood by analysis of the thermodynamic transition levels, whereas the deep nature of the electronic features observed in spectroscopies is rationalized on the basis of the computed optical transition levels. Our findings are in qualitative, and sometimes quantitative, agreement with previous investigations at the hybrid-functional\cite{deak2012,divalentin2014} and DFT+U\cite{mattioli2010} level.

\begin{figure}[tb]
 \includegraphics{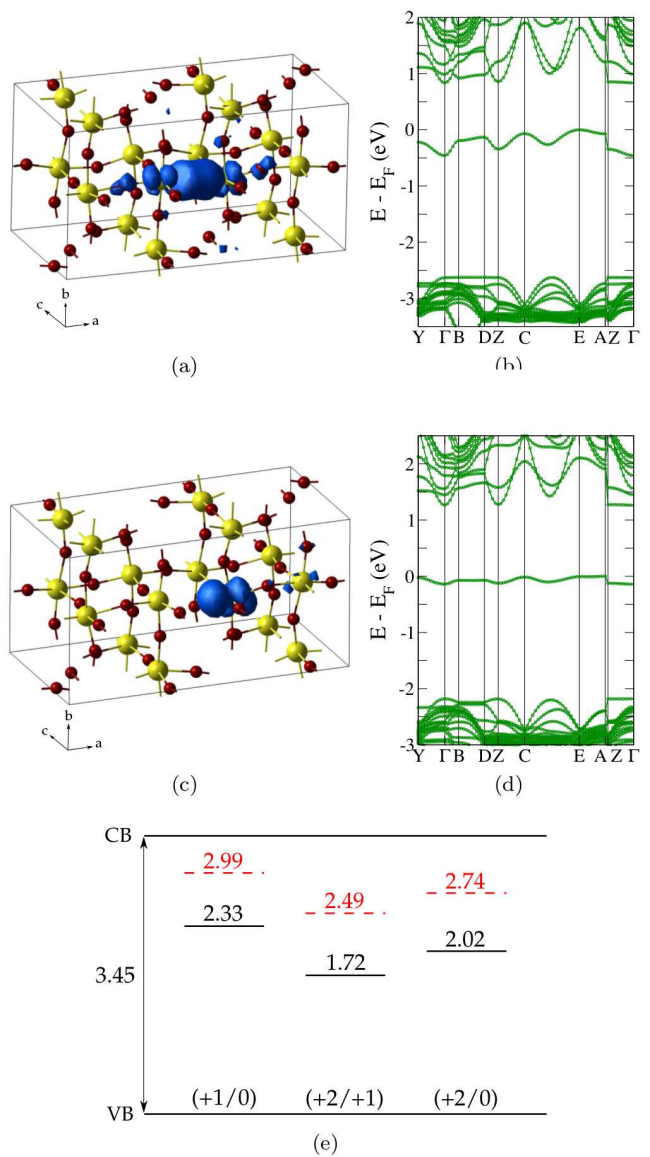}
 \caption{O vacancy in RT monoclinic WO$_3$. Isosurface plot of the charge density associated with the excess electrons for the (a) neutral and (c) singly-charged vacancy. The yellow (large) and red (small) spheres represent W and O atoms, respectively. Corresponding electronic band structures, computed at the equilibrium geometry, for the (b) neutral and (d) singly-charged vacancy (only spin majority is shown). (e) Optical (solid, black) and thermodynamic (dashed, red) charge-transition levels. Positions (in eV) are given with respect to the top of the VB.}
 \label{fig_3}
\end{figure}

\subsection{WO$_3$}

The optical and electrical properties of substoichiometric WO$_3$ (WO$_{3-x}$) have been extensively investigated experimentally, particularly in relationship with the electrochromism exhibited by this material.\cite{deb1973} However, the microscopic origin of this effect, which is intimately connected with the typical blue coloration of WO$_{3-x}$ films,\cite{deneuville1978,hollinger1976} has been longly debated, and several models have been put forward for rationalizing experimental observations.\cite{deb1973,faughnan1975,deb2000}

On the theory side, a few first-principle investigations have been performed on O-deficient RT monoclinic WO$_3$,\cite{lambert2006,chatten2005,migas2010} but among them only one\cite{wang2011a} employed state-of-the-art methods yielding a correct description of the bulk electronic structure. The latter study of Wang~\emph{et al.}\cite{wang2011a} revealed a delicate interplay between the concentration of O vacancies and the metallic or insulating nature of WO$_{3-x}$. Furthermore, different behaviors were observed according to the orientation of the W-O-W chain along which the O atom is removed. For the sake of simplicity, we will limit the present discussion to the case of an O vacancy created along the $a$ crystallographic axis of the monoclinic cell,\cite{note8} which we have evidence to believe to be the main responsible for the observed optical features of WO$_{3-x}$, namely its blue coloration. A more comprehensive investigation will be subject of a future publication.

Our calculations indicate the closed-shell singlet solution to be more stable (by 0.17~eV) than the open-shell triplet one for the system with a neutral vacancy. Correspondingly, the two extra electrons occupy a defect state located at about $1.0-1.5$~eV below the CB [Figure~\ref{fig_3}(a)]; they localize mainly at the vacancy void, with a contribution from the two under-coordinated W atoms [Figure~\ref{fig_3}(b)].

As the vacancy is ionized, the extra electron localizes on one of the under-coordinated W atoms [Figure~\ref{fig_3}(c)], leading to the formation of a reduced W$^{5+}$ ionic state. In fact, one of the proposed models for the chromic mechanism in WO$_{3-x}$ suggests that the presence of W$^{5+}$ centers results from the localization of excess electrons on W$^{6+}$ sites upon light absorption.\cite{faughnan1975} As shown in Figure~\ref{fig_3}(d), this $5d$ electron occupy a defect state $\sim 1.3$~eV below the CB ($\sim 2$~eV if the geometry of the neutral vacancy is retained).

Several experiments have been performed to elucidate the spectroscopic properties of substoichiometric WO$_3$ samples. The coloration efficiency of WO$_{3-x}$ ($x<0.4$) films has been reported to increase with increasing O deficiency.\cite{bechinger1997} Their typical blue coloration has been correlated with a broad absorption band with a maximum at 900~nm ($\sim 1.38$~eV),\cite{hollinger1976,deneuville1978} although this feature has been observed in amorphous films. Photoluminiscence (PL) spectra\cite{paracchini1982} exhibited an emission peak at 550~nm ($\sim 2.26$~eV), and photoelectron measurements\cite{hollinger1976,brigans1981} confirmed the presence of a defect state attributed to O vacancies at $\sim 2$~eV above the VB. 

The computed optical transition levels reported in Figure~\ref{fig_3}(e) are consistent with these observations, being positioned in the band gap in the range from 1.72~eV to 2.33~eV above the VB. The dispersion of the neutral-vacancy defect state limits the accuracy of the calculated transition levels to $\sim 0.2$~eV. Notice that the computed band gap is larger than the reported values from absorption measurements (see Ref.~\onlinecite{wang2011} and references therein), but in excellent agreement with ultraviolet direct/inverse photoemission experiments (3.45~eV from Ref.~\onlinecite{meyer2010}), as well as with $GW$ calculations.\cite{gerosa2015,ping2013} Thus, the excitation energy pertaining to, e.g., the $(+1/0)$ transition, which is obtained to be $1.12$~eV, is compatible with the experimentally observed features in the absorption spectra.

The corresponding thermodynamic levels are predicted to be well-detached from the bottom of the CB, with a minimum excitation energy of $0.46$~eV computed for the $(+1/0)$ transition. This implies that the specific kind of O vacancy considered here cannot contribute to the observed $n$-type conductivity of WO$_{3-x}$ films, which has been attributed to O vacancies.\cite{gillet2003} Thus, although the present work clarifies the nature of the spectroscopic features commonly evidenced in reduced WO$_3$, in agreement with the previous investigation of Wang~\emph{et al.},\cite{wang2011a} further study is called for to elucidate the possibly different behavior of other inequivalent O vacancy sites.

\begin{table}[tb]
\caption{\label{tab_1} Formation energy for the O vacancy in the different materials, computed with respect to $(1/2)E[\text{O}_2]$, where $E[\text{O}_2]$ is the ground-state total energy of the triplet O$_2$ molecule.}
\begin{ruledtabular}
\begin{tabular}{lcc}
 & $x$ & Formation energy (eV)  \\
\hline
TiO$_{2-x}$ (rutile) & 1/24 &  5.2 \\
TiO$_{2-x}$ (anatase) & 1/32 & 4.7 \\
WO$_{3-x}$ & 1/16 & 5.3  \\
ZrO$_{2-x}$ & 1/36 & 6.6 \\
\end{tabular}
\end{ruledtabular}
\end{table}

\begin{figure*}[tb]
 \includegraphics{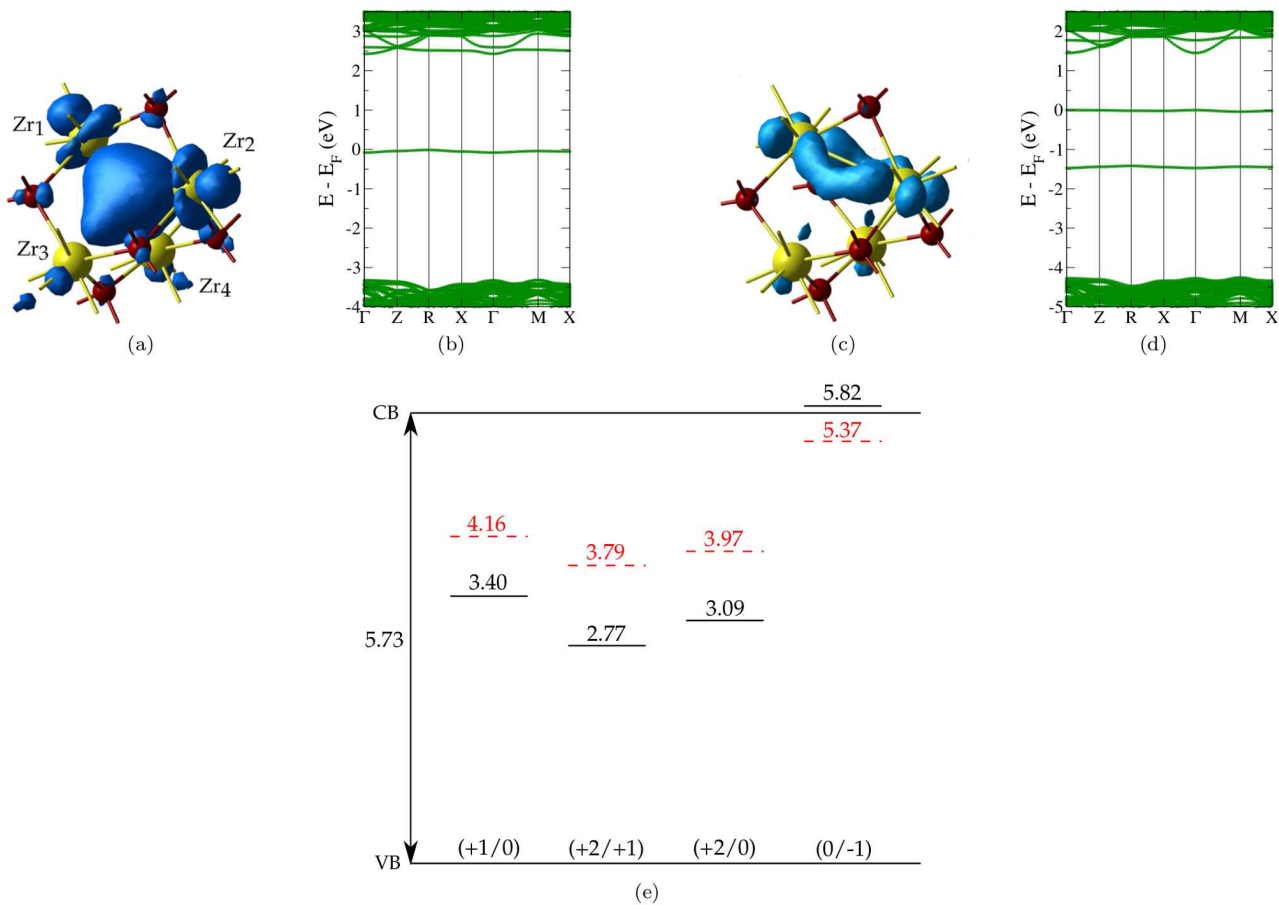}
 \caption{O vacancy in tetragonal ZrO$_2$. Isosurface plot of the charge density associated with (a) the two excess electrons for the neutral vacancy, and (c) the trapped electron for the negatively-charged vacancy. The yellow (large) and red (small) spheres represent Zr and O atoms, respectively. Corresponding electronic band structures, computed at the equilibrium geometry, for (b) the neutral, and (d) the negatively-charged vacancy (only spin majority is shown). (e) Optical (solid, black) and thermodynamic (dashed, red) charge-transition levels. Positions (in eV) are given with respect to the top of the VB.}
 \label{fig_4}
\end{figure*}

\subsection{ZrO$_2$}

The O vacancy in reduced zirconia (ZrO$_{2-x}$) has been reported to exhibit features similar to those of the F centers typically observed in non-reducible oxides,\cite{pacchioni2015} the extra charge being trapped in the vacancy void.\cite{gionco2013} EPR studies\cite{gionco2013} indicate that ZrO$_2$ is generally less prone to lose oxygen than other reducible oxides, such as TiO$_2$. This is confirmed by our calculations, giving a formation energy for the O vacancy in ZrO$_{2-x}$ larger by more than 1~eV with respect to the other oxides (see Table~\ref{tab_1}), in agreement with previous investigations.\cite{ganduglia2007,wang2011a}

As shown in Figure~\ref{fig_4}(a), upon removal of an O atom in ZrO$_2$, the two excess electrons stabilize in the vacancy void, with a contribution from the $4d$ orbitals of two nearest-neighbor Zr atoms (Zr$_1$ and Zr$_2$); the corresponding ground state is a closed-shell singlet. As expected, the neighboring Zr atoms relax towards the vacancy, resulting in a shrinking of the distance Zr$_1$-Zr$_2$ from 3.61~$\text{\AA}$ to 3.48~$\text{\AA}$, and of Zr$_1$-Zr$_3$ from 3.65~$\text{\AA}$ to 3.61~$\text{\AA}$. The corresponding band structure [Figure~\ref{fig_4}(b)] exhibits a flat defect state positioned at $\sim 3.3$~eV above the VB, in close agreement with a previous B3LYP study.\cite{gionco2013}

As the vacancy is ionized, the remaining extra electron distributes on the two Zr$_1$ and Zr$_2$ atoms, as well as in the space between them, in a typical bonding configuration. The extra electron occupies a doublet state in the band gap at $\sim 3.6$~eV above the VB.

We also studied the case of the negatively-charged O vacancy, which can be realized as a consequence of the trapping of an electron from the bulk CB. Such trapping phenomena can affect, for instance, the performance of electronic devices, for which high-$\kappa$ oxide materials such as zirconia or hafnia (HfO$_2$) have been proposed as gate electrodes.\cite{robertson2004} The trapped electron occupies an additional defect state in the band gap at $\sim 1.2$~eV above the (now spin-resolved, with spin majority shown) states associated with the neutral vacancy [Figure~\ref{fig_4}(d)], in a bonding configuration between the Zr$_1$ and the Zr$_2$ atoms [Figure~\ref{fig_4}(c)].

The computed CTLs corresponding to both vacancy ionization and electron trapping are shown in Figure~\ref{fig_4}(e). A band gap of 5.73~eV is calculated, to be compared with the experimental value of 5.78~eV deduced from optical absorption measurements;\cite{french1994} typically $GW$ calculations give a band gap larger by $\sim 0.1-0.2$~eV (see Ref.~\onlinecite{gerosa2015} and references therein).

Much experimental work is available in the literature investigating the PL properties of ZrO$_{2-x}$. However, a clear understanding of the underlying mechanisms seems to be lacking. Several emission features in the region between 2.0 and 3.5~eV are typically observed, and have been related to the presence of O vacancies.\cite{smits2010} The large uncertainty in experimental results may be the consequence of the practical difficulty in obtaining high-purity single crystals of zirconia with a well-defined phase.

PL measurements on tetragonal nanocrystalline ZrO$_{2-x}$ showed an emission peak at 350~nm ($\sim 3.54$~eV), which was attributed to electron-hole recombination involving F-center states in the band gap. Analysis of the PL spectrum evolution with annealing allowed to assign this feature to O vacancies.\cite{cong2009} Our computed position of the $(+1/0)$ optical transition level (3.40~eV above the top of the VB) may account for this observation.

Another study on undoped tetragonal zirconia nanocrystals\cite{smits2007} evidenced the presence of an emission peak at $\sim 2.8$~eV. This may be related to a PL process involving the $(+2/+1)$ optical level, which we found at 2.77~eV above the top of the VB.

The thermodynamic levels are found below the CB by at least 1.5~eV, indicating that all the O-vacancy charged states are stable at room temperature. An exception is represented by the $(0/-1)$ level, which is separated from the CB by $\sim 0.3$~eV. This level is associated to electron trapping processes causing the O-vacancy to become negatively-charged. Our calculations suggest that such trapping center is stable at room temperature, and the same conclusion was obtained from a similar investigation on hafnia\cite{broqvist2006} in which, due to analogous chemical behavior of the Hf and Zr cations, O vacancies are expected to induce similar effects as in zirconia.\cite{zheng2007} In this respect, trapping/detrapping experiments on HfO$_2$ found the activation energy for the trap level to be 0.35~eV,\cite{ribes2006} which is in agreement with our calculations.

\section{Conclusions}
\label{sec:conclusions}

We reviewed the behavior of the O vacancy in prototypical reducible (TiO$_2$, WO$_3$) and non-reducible (ZrO$_2$) oxide semiconductors adopting a novel DFT computational approach based on hybrid functionals with the exact-exchange fraction obtained \emph{ab initio} for the bulk non-defective material. Its capability of yielding accurate electronic band structures, together with structural properties and total energies, makes it an ideal tool for investigating defect levels in semiconductors.

The computed optical transition levels are generally found in agreement with various spectroscopy experiments, providing support for their interpretation. Instead, thermodynamic levels allow one to infer about the intrinsic degree of $n$-type conductivity of substoichiometric oxide materials through analysis of the stability of the different O-vacancy charge states. Residual discrepancies between theory and experiment may arise from two main sources: (i) the always present approximate treatment of xc terms, and (ii) usage of finite supercells. While the former is intrinsically related to the adopted functional, the latter may show up also as consequence of the evaluation of CTLs via the Janak's theorem, Eq.~\eqref{eq:janak-slater}, instead of by direct computation of total energy differences,\cite{note9} due to the defect band dispersion introduced by the residual polaron overlapping in neighboring cells.

Fundamental differences are evidenced in the behavior of reducible and non-reducible oxides. Upon creation of an O vacancy, the former tend to exhibit polaron-like features: the excess charge mainly localizes on the metallic ions, reducing their formal charge, and the relaxation of the atomic environment is not restricted to the atoms in the immediate vicinity of the defect. Instead, in non-reducible oxides such as ZrO$_2$, the excess charge is stabilized in the vacancy void, and the relaxation is more local; furthermore, transition levels are deeper, being situated near the center of the band gap. These are the fingerprints of the F-center-like behavior of O vacancies in these materials.

In conclusion, we believe that the present approach is able to characterize defect levels in semiconductors in a way that makes it valuable for predictive studies also on extrinsically-doped semiconductors. Notice however that, while the method is straightforwardly applied to point (zero-dimensional) defects, its extension to the treatment of higher-dimensional defect structures, such as surfaces, requires facing unprecedented difficulties. For instance, modeling the interface between two different semiconductors implies the exchange fraction to become a space-dependent quantity, and this eventually calls for a phenomenological model of the optical properties of the interface. To the best of the author's knowledge, the issue has not yet been considered in the literature, and its solution would make it possible to apply the method to investigate the rich physical-chemical phenomena occurring at interfaces.

\begin{acknowledgments}

This work has been supported by the Italian MIUR through the FIRB Project RBAP115AYN ``Oxides at the nanoscale: multifunctionality and applications''.
The support of the COST Action CM1104 “Reducible oxide chemistry, structure and functions” is also gratefully acknowledged. G.O. and L.C. acknowledge
the ETSF-Italy\cite{etsf} for computational support.

\end{acknowledgments}

\providecommand{\noopsort}[1]{}\providecommand{\singleletter}[1]{#1}%


\begin{thebibliography}{98}
\bibitem [{note1()}]{note1}%
  \BibitemOpen
  \bibinfo {note} {The same procedure will be understood in the following where
  not otherwise stated.}\BibitemShut {Stop}%
\bibitem [{note2()}]{note2}%
  \BibitemOpen
  \bibinfo {note} {The present interpretation of Eq.~\protect \textup {\hbox
  {\mathsurround \z@ \protect \normalfont (\ignorespaces \ref
  {eq:janak-slater}\unskip \@@italiccorr )}} is apt to donor states; for the
  dual case of acceptor states, an analogous interpretation in terms of KS hole
  states is easily figured out.}\BibitemShut {Stop}%
\bibitem [{note3()}]{note3}%
  \BibitemOpen
  \bibinfo {note} {Notice that usage of the total dielectric constant
  (including the ionic contribution) is improper in this case, since in optical
  excitations ions are assumed to be fixed, and thus cannot contribute to
  screening of the additional defect charge arising, for instance, in the
  $(q/q+1)$ transition. Only the electronic subsystem will be polarized by it,
  hence the usage of $\epsilon _{\infty }$.}\BibitemShut {NoStop}%
\bibitem [{note5()}]{note5}%
  \BibitemOpen
  \bibinfo {note} {Such solution was obtained by augmenting the basis set with
  Gaussian-type orbitals corresponding to the ghost O atom at the vacancy;
  furthermore, a polaron-like configuration was stabilized using the very same
  basis set, which allowed us to meaningfully compare the corresponding total
  energies.}\BibitemShut {Stop}%
\bibitem [{note6()}]{note6}%
  \BibitemOpen
  \bibinfo {note} {We have also been able to stabilize a different triplet
  solution in which the point-group symmetry is preserved; however we found it
  to be 0.23~eV higher in energy than the symmetry-broken one.}\BibitemShut
  {Stop}%
\bibitem [{note8()}]{note8}%
  \BibitemOpen
  \bibinfo {note} {Along the W-O-W $a$-chain, W-O bonds throughout have
  practically the same lengths, while for chains along the $b$ and $c$ axis,
  alternating long and short bonds are present; this peculiar structural
  feature possibly determines a different behavior of the O
  vacancy.}\BibitemShut {Stop}%
\bibitem [{note9()}]{note9}%
  \BibitemOpen
  \bibinfo {note} {Such differences cannot be computed with the \textsc{crystal09} code.}\BibitemShut {Stop}%
\end{thebibliography}
\end{document}